\documentclass[a4paper,fleqn,usenatbib]{mnras}

\usepackage{newtxtext,newtxmath}

\usepackage[T1]{fontenc}
\usepackage{ae,aecompl}

\usepackage{graphicx}	
\usepackage{amsmath}	
\usepackage{amssymb}	

\title[Effects of albedo and disc on the zero velocity \dots]{Effects of albedo and disc on the zero velocity curves and  linear stability of equilibrium points in the generalized restricted three body problem}

\author[Saleem Yousuf and Ram Kishor]{
Saleem Yousuf\thanks{E-mail: sf07bhu@gmail.com}
Ram Kishor\thanks{E-mail: kishor.ram888gmail.com}
\\
 Department of Mathematics, Central University of Rajasthan, NH-8, Bandarsindari, Kishangarh, Ajmer-305817, Rajasthan, India\\
}

\date{Accepted XXX. Received YYY; in original form ZZZ}

\pubyear{2015}

\begin{document}
\label{firstpage}
\pagerange{\pageref{firstpage}--\pageref{lastpage}}
\maketitle

\begin{abstract}
The most important aspects of a dynamical system are its stability and the factors which affects the stability property. This paper presents the analysis of the effects of albedo and disc on the zero velocity curves, existence of equilibrium points and on their linear stability in a generalized restricted three body problem that consists of motion of an infinitesimal mass under the uniform gravity field of radiating-oblate primary, oblate secondary and a disc, which is rotating about the common center of the mass of the system. A significant effect of albedo and disc are observed on the zero velocity curves, positions of equilibrium points and on the stability region. Linear stability analysis of collinear equilibrium points is performed with respect to mass ratio $\mu$ and albedo parameter of secondary, separately and it is found that these are unstable in both the cases. On the other hand, non-collinear equilibrium point is stable in a certain range of mass ratio. After analyzing individual as well as combined effect of radiation pressure force of the primary, albedo of secondary, oblateness of both the massive bodies and the disc, it is found that these perturbations play a significant on the motion of infinitesimal mass in the vicinity of equilibrium points. These results may be help to analyze more generalized problem of few bodies under the influence of different kind of perturbations such as P-R drag, solar wind drag etc. Present study is limited to the regular symmetric disc which will  extend later. 
\end{abstract}




\section{Introduction}
\label{sec:intro}

The restricted three body problem (RTBP) has been a continuous source of study from last few centuries. In the RTBP, two larger masses move in a circular orbit about the center of their masses with uniform speed and third body, which is of negligible mass also known as infinitesimal mass, moves under the gravitational attraction of two larger masses, without influencing them. The study of equilibrium points in RTBP has a great importance for the astronomers and physicists in the context of analysis of different space missions. For example $L_1$ point is home to Solar and Heliospheric Observatory (SOHO) satellite and $L_2$ point is home to Wilkinson Microwave Anisotropy Probe (WMAP) spacecraft and James Webb Space Telescope, respectively, whereas $L_{4,5}$ of the Sun-Jupiter system are home to Trojan asteroids. Still, RTBP is an active and motivational research field which has been attracting large number of scientists, astronomers and researchers. In classical RTBP, there exist five equilibrium points denoted as $L_{1}$, $L_{2}$, $L_{3}$, $L_{4}$ and $L_{5}$. Three of them i.e. $L_{1}$, $L_{2}$ and $L_{3}$ are known as collinear equilibrium points, which are unstable in the range $0<\mu\leq\frac{1}{2}$ and two of them i.e. $L_{4}$, $L_{5}$ are called non-collinear equilibrium points, which are stable in the range $0<\mu<\mu_{c}=0.0385209$ \citep{Szebehely1967torp.book.....S}.  

Because of continuous radiation  consequently radiation pressure force from the Sun, the gravitational attraction force due to the Sun is reduced in comparison to that of its actual magnitude, which affects the classical results. \cite{Simmons1985CeMec..35..145S} have considered both primaries as luminous body and described complete existence and linear stability of the equilibrium points for all values of mass ratio as well as radiation pressure of both the luminous bodies. They also found four additional out of plane equilibrium points $L_{6}$, $L_{7}$, $L_{8}$ and $L_{9}$. Many researchers \citep{Chernikov1970SvA....14..176C,meyer1986stability,Liou1995Icar..116..186L,Ragos1995A&A...300..568R,haque1995non,gozdziewski1998nonlinear,abouelmagdshaboury2012Ap&SS.341..331A,singhtaura2013Ap&SS.343...95S}  have discussed the dynamical behavior in the photo-gravitational restricted three body problem. The effect of radiation pressure and PR-Drag are examined by \citep{das2008Ap&SS.314..261D} during the study of motion of dust particles in the steller binary systems: RW-Monocerotis and Kr{\"u}ger 60.
As, the Sun is the source of radiation in the solar system and the radiations from the Sun is incident on the surface of the Earth from which some part is reflected back in space. This phenomenon of solar radiation reflecting back in space is known as albedo. It is a dimensionless quantity and also defined as the measure of reflectivity of the planet's surface \citep{harris1969albedo}. The albedo varies in $[0,\,1]$. On an average, Earth reflects about $30\%$ of total incident radiations of Sun's energy i.e Earth's albedo is nearly $0.3$.
\cite{Anselmo1983A&A...117....3A,bhanderi2005modeling} have analyzed the effect of Earth's albedo on the Earth bound satellites.  Few researcher \citep{mcinnes1994solar,abdel2011new,gong2015analytical,Idrisi2017JAnSc..64..379I} have studied the RTBP under albedo effect of secondary in addition to the oblateness effect. \cite{genio2017} studied the predicted temperature and predicted albedo for the exo-planets such as Kepler-186 f, Proxima Centauri b, etc. Recently, \cite{idrisi2018non} have discussed the elliptic restricted three body problem under the influence of albedo and they found that positions of $L_{4,5}$ get affected, significantly. Thus, albedo effect have a significant impact on the natural or artificial satellites in the planetary systems and may be considered as perturbation factor in scientific missions or steller binary systems. 

Due to lack of sphericity of the planets such as Earth, Saturn, Jupiter etc., which are sufficiently oblate spheroid and tri-axial ellipsoid, the oblateness and triaxiality of the massive body is an other important factor to be considered in the real problem as a perturbation, which influence the system. \cite{Subbarao1975A&A....43..381S} have studied the problem for non-collinear equilibrium points considering both primaries as oblate spheroid and found that oblateness decreases the range of stability. Many authors \citep{Markellos1996Ap&SS.245..157M,Ishwar2001Ap&SS.277..437I,Singh2014JApA...35..729S,Zotos2015Ap&SS.358...33Z,Singh2016EPJP..131..365S,singhrichard2017EPJP..132..330S,singhumar2017DEDS...25...11S} have analyzed the stability of equilibrium points in restricted three body problem under the influence of oblateness and triaxiality of the primaries.  

In recent years, there have made new discoveries of dust belts around the stars with their known planets \citep{Greaves2004MNRAS.351L..54G,Jayawardhana2000ApJ...536..425J,Chavez2016MNRAS.462.2285C}. \cite{anglada2017alma} found the dust belts in the Proxima Centauri system, which is analog to Kuiper belt in the solar system. The presence of disc or belt like structure such as Kuiper belt, asteroid belt in our solar system, disc of dusts in extra-solar planetary systems etc., affect the dynamical properties of the planetary systems. \cite{jiang2001planetary} investigated the orbital elements with an interaction between exoplanets and the  proto-steller disc, in the  planetary system of upsilon Andromedae. The influence of asteroid belt in between Mars and Jupiter in the solar system was observed by \cite{jiang2003bifurcation}. He found that probability of equilibrium points is larger near the inner part of the belt than outer one.
\cite{jiang2004chaotic,jiang2004modified,jiang2004dynamical,jiang2006chermnykh} studied the orbital behavior and the stability of equilibrium points in the disk-star-planet model, which plays a significant role in the Kirkwood gaps formation in our solar system. In addition to $L_{1,2,3,4,5}$, they have found new collinear equilibrium points JY1 and JY2, known as Jiang-Yeh points, which exist only if the total mass of the galaxy is greater than that of critical mass. Several researchers  \citep{Kushvah2008Ap&SS.318...41K,Kushvah2012Ap&SS.337..115K,kishor2013linear,falaye2016triangular} have also studied the linear stability of equilibrium points of the restricted problem of three bodies with the inclusion of disc-like structure.
Recently, \cite{jiang2014galaxy1,jiang2016galaxy3} modified the RTBP by taking a star as a test particle, moving in the gravitational influence of super-massive binary black hole and the external galactic potential.  

Interplanetary dusts, comets, asteroids etc., which seem very small, forms a disc  like structure in space such as HD 95086 binary system with two dust belt, Proxima Centauri system with a dust belt etc. These massive disc like structure may change the dynamical behavior of infinitesimal mass and hence, its orbit. \cite{idrisi2018non} have described the RTBP with oblateness and albedo of the Earth during the study of motion of Earth's artificial satellite. Thus, it is reasonable to develop a model, which may study the effect of albedo as well as potential influence of the disc on the equilibrium points and their linear stability. Motivating from the above arguments and their importance in the realistic problem, we are interested to analyze the effect of albedo in the context of equilibrium points and their linear stability in the Proxima Centauri system under the frame of RTBP in the presence of radiation pressure due to radiating primary (Proxima Centauri), oblateness of the primaries (Proxima Centauri and Proxima Centauri b) and a disc (dust belt). In spite of these, we also have obtained the albedo effect in Sun-Mars-asteroid belt and Sun-Saturn-Kuiper belt.
The present work may be helpful in the study of the motion of satellites, space probes such as New Horizons, Poineer 10, Poineer 11, Voyager 1 and Voyager 2 etc., which are specificially designed to explore the outer planets, KBO's, launched by different space agencies such as NASA, ESA etc.

The contents of the paper are organized as follows: Section-\ref{sec:eqnmotion} deals the formulation of the problem. The analysis of zero velocity curves is presented in Section-\ref{zvc}.  Section-\ref{sec:eqmpts} contains the computations of equilibrium points under the effect of perturbations, whereas Section-\ref{sec:linearstab} reflects the linear stability analysis of the equilibrium points in addition to the computation of perturbed mass ratio of the problem. Finally, paper is concluded in  Section-\ref{sec:conclusion}.  Numerical as well as some of the algebraic computation are performed with the help of latest version of Mathematica software. 
\section{Equations of motion}
\label{sec:eqnmotion}
\begin{figure}
	\includegraphics[width=0.45\textwidth]{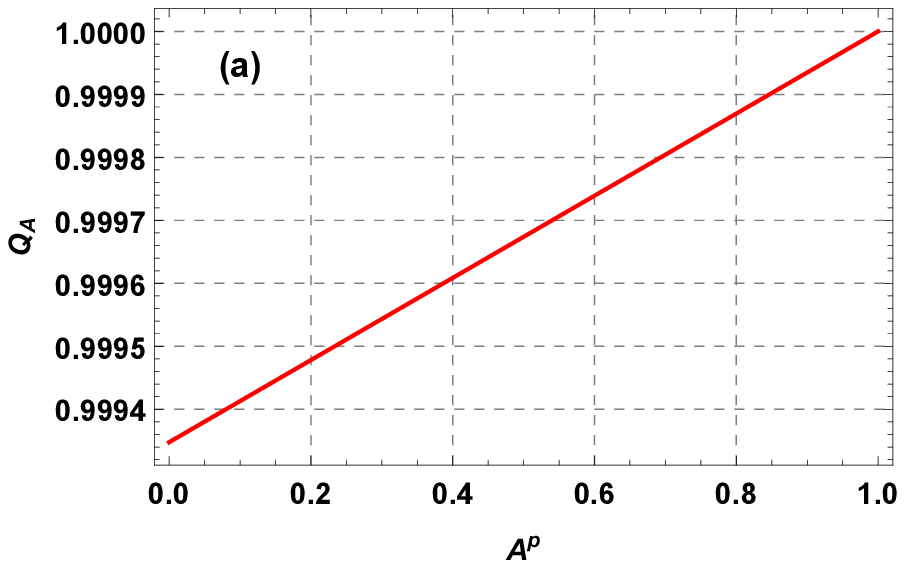} \\
	\includegraphics[width=0.45\textwidth]{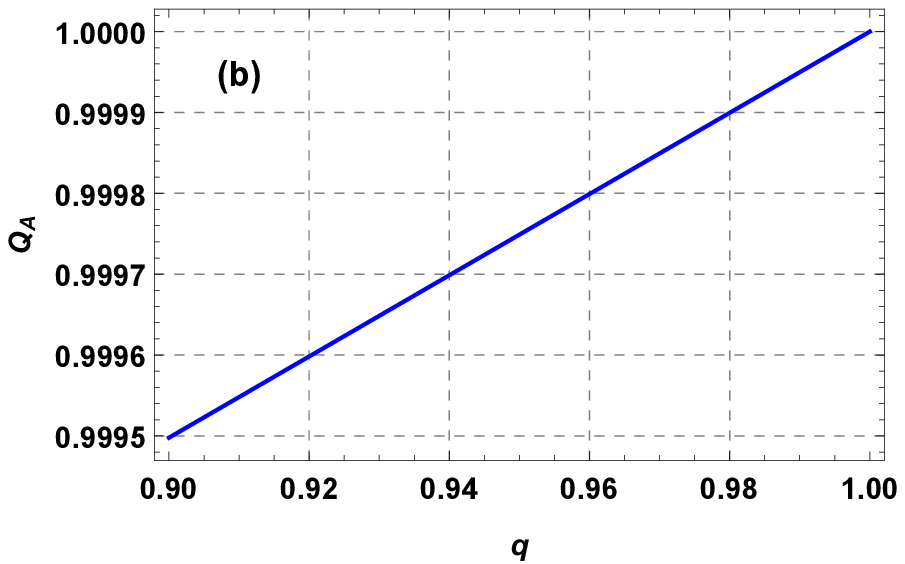} 
	\caption{Variation of albedo parameter $Q_{A}$ in Proxima Centauri  system at $\mu$=$0.000031$ for (a) $q=0.90$ and (b) $A^{p}=0.23$.}\label{fig:albedo}
\end{figure}

Consider a perturbed RTBP,  which consists of motion of an infinitesimal mass under the influence of uniform gravity field of two massive bodies and a disc.   The first body (bigger primary) with mass $m_{1}$ is a radiating oblate body, second body (smaller primary) with mass $m_{2}<m_{1}$ is only oblate spheroid and the disc with mass $M$  is rotating about the common center of mass of the system.   Let, the distance between the primaries and the sum of their masses be the units of distance and mass, respectively and the unit of time be the time period of the system so that $G(m_1+m_2)=1$. Assume that the effect of gravitational field of the infinitesimal mass on the remaining system is negligible.  
Let $F_{1}$ and $F_{2}$ be the gravitational forces, which are acting on infinitesimal mass due to $m_{1}$ and $m_{2}$, respectively. Let $F_{p}$  be the  radiation pressure force due to $m_{1}$ and $F_A$ be albedo force due to $m_2$, which are  acting on the infinitesimal mass towards the directions opposite to the directions of $F_1$ and $F_2$, respectively.  
Therefore, the resultant force acting on infinitesimal mass due to the first and second primary are \citep{Idrisi2017JAnSc..64..379I,idrisi2018non}
\begin{eqnarray}
q F_{1}=F_{1}\left(1-\frac{F_{p}}{F_{1}}\right)=F_{1}(1-\sigma_{1}),\\
\text{and}\quad Q_A F_{2}=F_{2}\left(1-\frac{F_{A}}{F_{2}}\right)=F_{2}(1-\sigma_{2}),
\end{eqnarray}
respectively, where $q$ is mass reduction factor of first primary and $Q_{A}$ is albedo parameter of second primary, which are defined as  $q=1-\sigma_{1}$, $Q_{A}=1-\sigma_{2}$ with $\sigma_{1}=\frac{F_{p}}{F_{1}}<<1$ and $\sigma_{2}=\frac{F_{A}}{F_{2}}<<1$. Also, $\sigma_{1}=\frac{L_{1}}{2\pi G m_{1}c\kappa}$ and $\sigma_{2}=\frac{L_{2}}{2\pi Gm_{2}c\kappa}$ with $L_{1}$, $L_{2}$ as the luminosity of the first and second primary, respectively; $G$ is the gravitational constant; $c$ is the speed of light and $\kappa$ is the mass per unit area of the infinitesimal mass.
By using Stefan-Boltzmann law, the luminosity of primaries can be expressed as \citep{johnson2012mathematical}
$ L_{1}=4\pi R_{1}^{2}\sigma T_{e_{1}}^{4}$ and $L_{2}=4\pi R_{2}^{2}\sigma \epsilon T_{e_{2}}^{4},
$ where, $R_{1,2}$, $T_{e_{1},e_{2}}$ are the radius and effective temperature of the first and second primary, respectively; $\sigma$ is the Stefan-Boltzmann constant and $\epsilon$ is the emissivity of second primary. Now we introduce albedo of second primary by rewriting luminosity $L_{2}$ in terms of predicted albedo $A^{p}$ \citep{genio2017} as $L_{2}=\pi R_{2}^2\epsilon S_{0}(1-A^{p})$ where, $S_{0}$ is a solar constant. Therefore,
\begin{equation}
\frac{\sigma_{2}}{\sigma_{1}}=\left(\frac{1-\mu}{\mu}\right)(1-A^{p})k,\label{eqn:ratio1} 
\end{equation}
where $k=\frac{R_{2}^2\epsilon S_{0}(1-A^{p})}{4 R_{1}^{2}\sigma T_{e_{1}}^{4}}$ such that $0<k<<1$, $0\leq\sigma_{1}<<1$ and $0\leq\sigma_{2}<\sigma_{1}$. In terms of $q$ and $Q_{A}$, the above ratio reduces to
\begin{equation}
Q_{A}=1-\left(\frac{1-\mu}{\mu}\right)(1-q)(1-A^{p})k,\label{eqn:ratio2}
\end{equation}
where $0<q\leq1$, $q<Q_{A}\leq1$ and $A^{p}$ lies between 0 (if all incident radiation absorbed) and 1 (if all incident radiation reflected). Also, the value of $k$ depends on the system in question. The variation of albedo parameter $Q_{A}$  with the change in predicted albedo ($A^{p}$) and mass reduction factor (q) of the Proxima Centauri system are shown in Figure  \ref{fig:albedo}. From Figure \ref{fig:albedo}, it is clear that the range of $Q_A$ is $0.9<Q_A\le1$ for the Proxima Centauri system.
As the masses $m_1$ and $m_2$ move in the same plane under the mutual gravitational attraction so the potential $V_1$ between the primaries due the effect of oblateness can be expressed as \citep{murray1999,abouelmagd2012}
\begin{equation}
V_1 = -\frac{G\,m_1\,m_2}{R}\left[1+\frac{3(A_{21}+A_{22})}{2\,R^2}\right]
\end{equation} 
where, $A_{21}$ and $A_{22}$ are the non-dimensional oblateness coefficients of first and second primary, respectively, which are given as $A_{2i}=\dfrac{R_{ei}^2-R_{pi}^2}{5R^2},\,i=1,2$ \citep{mccuskey1963introduction}, where $R_{ei},\,R_{pi},\, i=1,2$ are  equilateral and polar radii of the primaries, respectively and $R$ is the distance between their centers.
Also, the potential from the dust belt \citep{miyamoto1975} written as 
\begin{equation}
V_2 = -\frac{G\,M_d}{(r_c^2+T^2)^{1/2}}
\end{equation}
where $M_d$ is the mass of dust belt; $r_c$ is the dimensionless reference radius of dust belt and  $T=a+b$ defines the density profile of the dust belt. 
Suppose  $l_1$ and $l_2$ be the distance of the first primary and second primary from their common center of mass.
Since, the primaries are assumed in circular motion with constant angular velocity  $n$ (mean motion), thus the motion of first and second primary along with the dust belt \citep{mccuskey1963introduction,Szebehely1967torp.book.....S,Kushvah2008Ap&SS.318...41K,singhleke2014Ap&SS.350..143S} can be written as 
\begin{eqnarray}
m_1\,n^2\,l_1=\frac{G\,m_1\,m_2}{R^2}\left[1+\frac{3(A_{21}+A_{22})}{2\,R^2}\right]+\frac{\,G\,M_d\,m_1\,r_c}{(r_c^2+T^2)^{3/2}}\label{eqnofm1}
\end{eqnarray}
and 
\begin{eqnarray}
m_2\,n^2\,l_2=\frac{G\,m_1\,m_2}{R^2}\left[1+\frac{3(A_{21}+A_{22})}{2\,R^2}\right]+\frac{\,G\,M_d\,m_2\,r_c}{(r_c^2+T^2)^{3/2}}\label{eqnofm2}.
\end{eqnarray}
Adding equation (\ref{eqnofm1}) and (\ref{eqnofm2}), we get 
\begin{eqnarray}
(l_1+l_2)n^2=\frac{G\,(m_1+m_2)}{R^3}\left[1+\frac{3(A_{21}+A_{22})}{2\,R^2}\right]+\frac{2\,G\,M_d\,r_c}{(r_c^2+T^2)^{3/2}}\label{eqnmm}.
\end{eqnarray}
If the distance between the primaries is $R=l_1+l_2=1$ unit, unit of mass is $m_1+m_2$ and unit of time is the time period of the frame so that $G(m_1+m_2)=1$. Thus, equation (\ref{eqnmm}) yields  
\begin{eqnarray}
n^2=\left[1+\frac{3(A_{21}+A_{22})}{2}\right]+\frac{2\,M_d\,r_c}{(r_c^2+T^2)^{3/2}}\label{eqnmm1},
\end{eqnarray}   
which is the expression for the mean motion.

Suppose, the co-ordinate of infinitesimal mass in the $xy$-plane is $(x,\,y,\,0)$ and that of the first and second primary are $(-\mu,\,0,\,0)$ and $(1-\mu,\,0,\,0)$, respectively, in synodic frame of reference $Oxyz$, where $\mu=\frac{m_{2}}{m_{1}+m_{2}}$ is the mass parameter. Then, equations of motion of the infinitesimal mass in the orbital plane \citep{Szebehely1967torp.book.....S,murray1999} are written as
\begin{equation}
\ddot{x}-2n\dot{y}=\frac{\partial \Omega}{\partial x},\label{eq:ox}
\end{equation}
\begin{equation}
\ddot{y}+2n\dot{x}=\frac{\partial \Omega}{\partial y},\label{eq:oy}
\end{equation}
with effective potential 
\begin{eqnarray}
&&\Omega=\frac{n^2}{2}(x^2+y^2)+\frac{q(1-\mu)}{r_{1}}\left(1+\frac{A_{21}}{2r_{1}^2}\right)\nonumber\\&&+\frac{Q_{A}\mu}{r_{2}}\left(1+\frac{A_{22}}{2r_{2}^2}\right)+\frac{M_{d}}{(r_{c}^2+T^2)^\frac{1}{2}},\label{eq:ep}
\end{eqnarray}
where $r=\sqrt{x^2+y^2}$, $r_{1}=\sqrt{(x+\mu)^2+y^2}$ and $r_{2}=\sqrt{(x+\mu-1)^2+y^2}$ are the distances of infinitesimal mass from center of mass, first primary and second primary, respectively and $n$ is the mean motion of the system which can be obtained from equation (\ref{eqnmm1}).
\section{Zero velocity curves}
\label{zvc}
The well known Jacobi integral of the problem in $xy$ plane is written as 
\begin{equation}
\dot{x}^2+\dot{y}^2=2\,\Omega-C_{j},
\end{equation}
where $\sqrt{\dot{x}^2+\dot{y}^2}$ is the velocity of the infinitesimal mass and $C_{j}$ is the Jacobi constant. As the velocity of the infinitesimal mass is zero at equilibrium points, hence the Jacobi Integral reduces to zero velocity curve as
\begin{equation}
C_{j}=2\,\Omega,\label{eq:jacobi}
\end{equation}
where equation (\ref{eq:jacobi}) defines a set of curves for the particular value of $C_{j}$  corresponding to respective equilibrium points $L_{i},\, i=1,2,3,4,5$. In restricted three body problem, the zero velocity curve permits to analyze the dynamical behavior of infinitesimal mass around the equilibrium points and determines the region in which infinitesimal mass is not allowed to move \citep{murray1999}.
 
\begin{figure*}
	\includegraphics[width=0.45\textwidth]{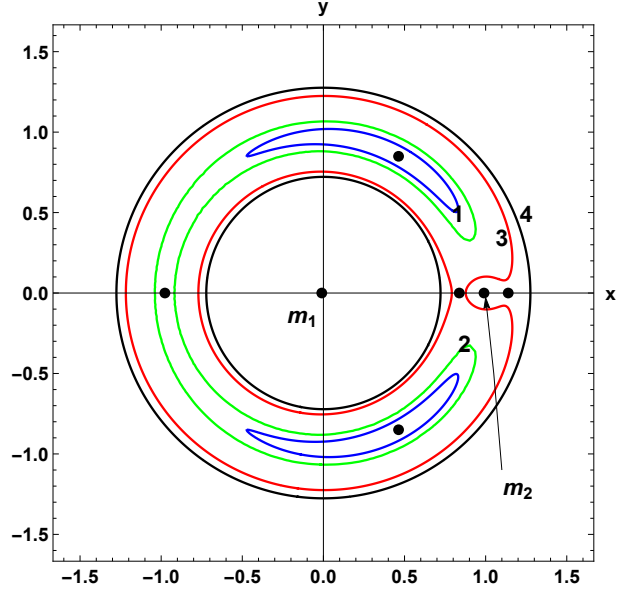} 
	\caption{Plot of the prohibited region for $\mu = 0.01$ at (\textbf{1}) $C_j=2.84$, (\textbf{2}) $C_j=2.86$, (\textbf{3}) $C_j=3.00001$ and (\textbf{4}) $C_j=3.07$.}\label{fig:zvca}
\end{figure*}

\begin{figure*}
	\includegraphics[width=0.50\textwidth]{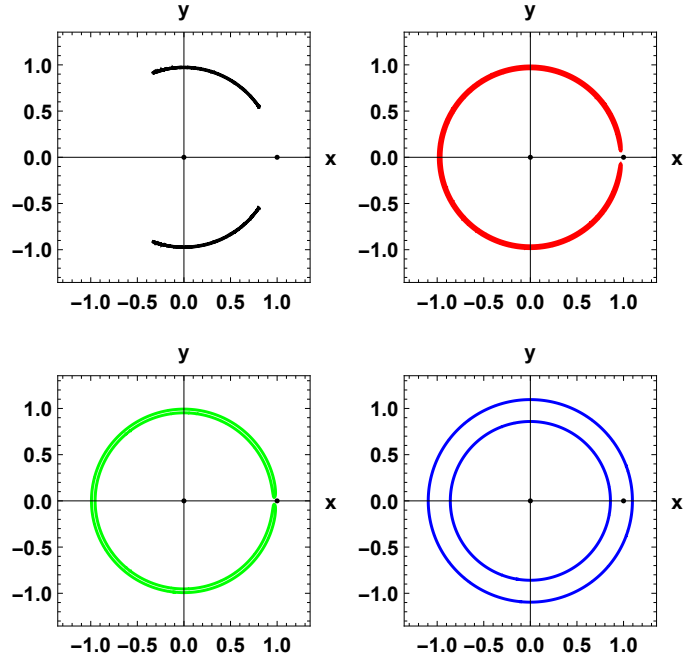} 
	\caption{Plot of the prohibited region at (\textbf{a}) $C_j=2.8378$, (\textbf{b}) $C_j=2.8385$, (\textbf{c}) $C_j=2.8390$ and (\textbf{d}) $C_j=3.01$ in the Proxima Centauri system with dust  belt.}\label{fig:zvcb}
\end{figure*}

\begin{figure*}
	\includegraphics[width=0.45\textwidth]{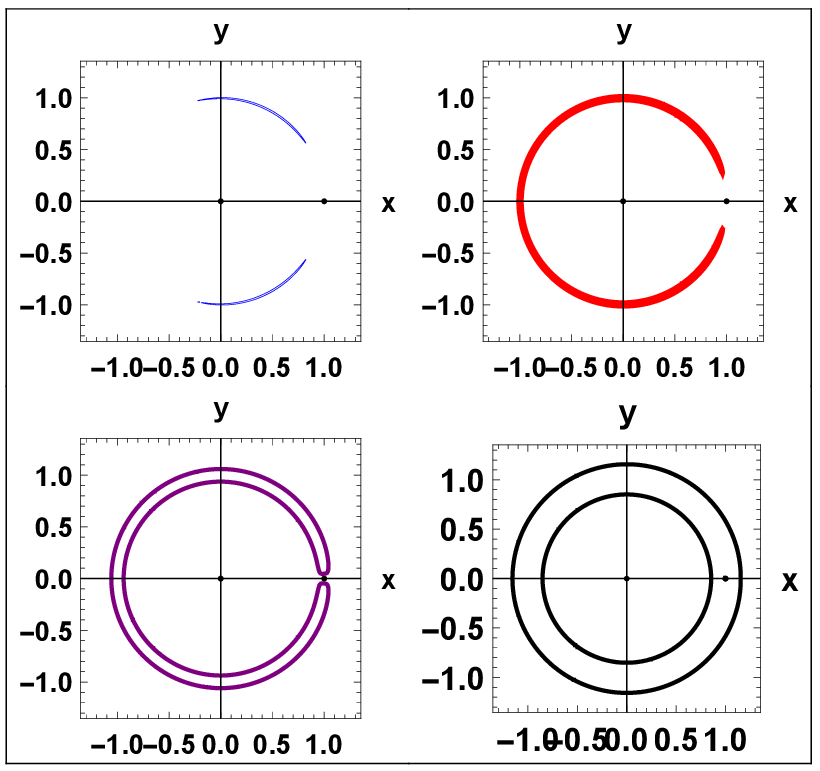} 
	\caption{Plot of the prohibited region at (\textbf{a}) $C_j=2.9799$, (\textbf{b}) $C_j=2.9810$, (\textbf{c}) $C_j=2.9910$ and (\textbf{d}) $C_j=3.05$ in the Sun-Saturn system with Kuiper belt.}\label{fig:zvcc}
\end{figure*}

 The zero velocity curves for the Proxima Centauri system with dust belt and Sun-Saturn with Kuiper belt are computed under the effects of radiation pressure $q$, albedo $Q_A$, oblateness of primaries $A_{21}$ and $A_{22}$, mass of the disc $M_{d}$ and these are show in Figs.  \ref{fig:zvca}-\ref{fig:zvcc}. Also, the values of $C_j$ under the influence of assumed  perturbations for different systems are presented in Table \ref{tabzvc}.
 \begin{table*}
 	\centering
 		\caption{Variation in the value of Jacobi constant $C_j: j=1,2,3,4,5$ in different systems.}\label{tabzvc}
 		\begin{tabular}{@{}|lcccc|@{}} 
 			\hline
 			Planetary System &$C_{L_{1}}$ &$C_{L_{2}}$ & $C_{L_{3}}$ & $C_{L_{4,5}}$\\
 			\hline
 			Proxima Centauri with dust belt & 2.83962 & 2.84705 & 2.83783 & 2.83847\\\hline
 			Sun-Mars with asteroid belt & 2.93975 & 2.94040 & 2.93970 & 2.93978\\\hline
 			Sun-Saturn with Kuiper belt & 2.99688 & 2.99833 & 2.98026 & 2.97970\\\hline
 		\end{tabular}
 \end{table*} 

From the Figs \ref{fig:zvca}-\ref{fig:zvcc}, it is observed that for low value of $C_j$, the restricted region for the infinitesimal mass lies around the equilibrium points $L_4$ and $L_5$. Further, on increasing the value of $C_j$, the prohibited region expand slowly then takes the form of a circular disc and engulfs the point $L_3$. For particular $C_j=3.01$ in Fig \ref{fig:zvcb}, the infinitesimal mass can move in a small region around the first primary $m_1$ and so infinitesimal mass are not allowed to move around  the second primary $m_2$. It is concluded that perturbation parameters affects the values of $C_j$  due to which results of zero velocity curves show the chaotic behavior.   

\section{Equilibrium point}
\label{sec:eqmpts}

Equilibrium point is a point at which motion of the test particle ceases. In the RTBP, equilibrium points are of two kind:  first, which lie on the line joining both the primaries, called collinear equilibrium point and second, which lie in the  orbital plane except collinear axis and known as non-collinear equilibrium point. The RTBP has three collinear equilibrium points, which are denoted as $L_1$, $L_2$ and $L_3$ and two non-collinear equilibrium points, which are denoted by $L_{4}$ and $L_{5}$. The non-collinear equilibrium point make an equilateral triangle with two vertices's at the primaries therefore, it is also known as triangular equilibrium point \citep{Szebehely1967torp.book.....S}. These equilibrium points can be obtained by solving $\dfrac{\partial \Omega}{\partial x}=0$ and $\dfrac{\partial \Omega}{\partial y}=0$ \citep{murray1999} i.e
\begin{eqnarray}
&&n^2x-\frac{q(1-\mu)(x+\mu)}{r_{1}^3}-\frac{Q_{A}\mu(x+\mu-1)}{r_{2}^3}-\nonumber\\&&\frac{3q(1-\mu)A_{21}(x+\mu)}{2r_{1}^5}-\frac{3Q_{A}\mu A_{22}(x+\mu-1)}{2r_{2}^5}-\nonumber\\&&\frac{M_{d}x}{(r_{c}^2+T^2)^{\frac{3}{2}}}=0,\label{eq:omx}
\end{eqnarray}
\begin{eqnarray}
&&n^2y-\frac{q(1-\mu)y}{r_{1}^3}-\frac{Q_{A}\mu y}{r_{2}^3}-\frac{3q(1-\mu)A_{21}y}{2r_{1}^5}
-\nonumber\\&&\frac{3Q_{A}\mu A_{22}y}{2r_{2}^5}-\frac{M_{d}y}{(r_{c}^2+T^2)^{\frac{3}{2}}}=0,\label{eq:omy}
\end{eqnarray}
for space variable $x$ and $y$.

\subsection{Collinear equilibrium points $(L_{1,2,3})$}
\label{sec:collinear}

Since, the collinear equilibrium point lies on the line joining the primaries on which $y=0$, consequently, equation (\ref{eq:omy}) becomes unimportant and  equation (\ref{eq:omx}) reduces to 
\begin{eqnarray}
&&f(x,0)=n^2x-\frac{q(1-\mu)(x+\mu)}{r_{1}^3}\left\{1+\frac{3A_{21}}{2r_{1}^2}\right\}\nonumber\\&&-\frac{Q_{A}\mu(x+\mu-1)}{r_{2}^3}\left\{1+\frac{3A_{22}}{2r_{2}^2}\right\}\nonumber\\&&-\frac{M_{d}x}{(x^2+T^2)^{\frac{3}{2}}}=0,\label{eq:cpt}
\end{eqnarray}
where $r_{1}=|x+\mu|$ and $r_{2}=|x+\mu-1|$. 
\begin{figure*}[h]
	\includegraphics[width=0.45\textwidth]{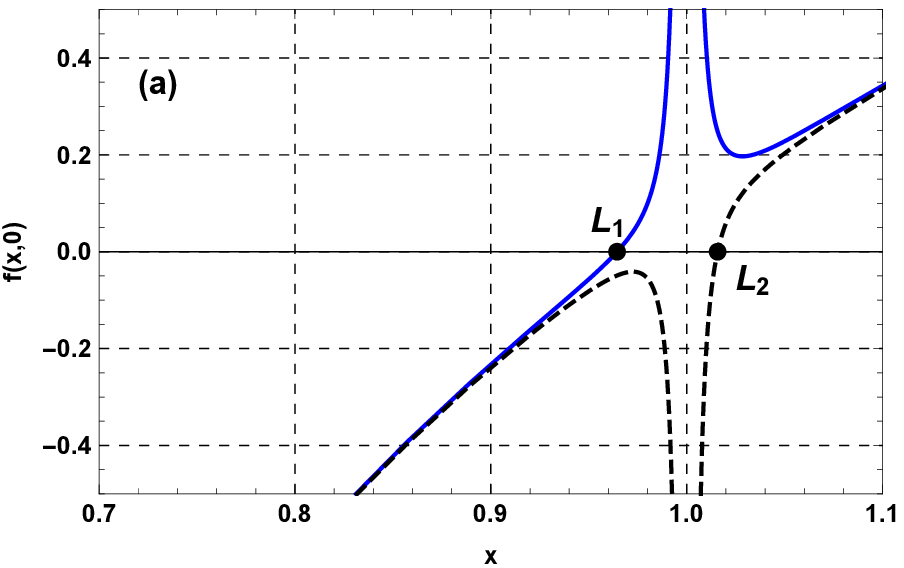} \hspace{2em}
	\includegraphics[width=0.45\textwidth]{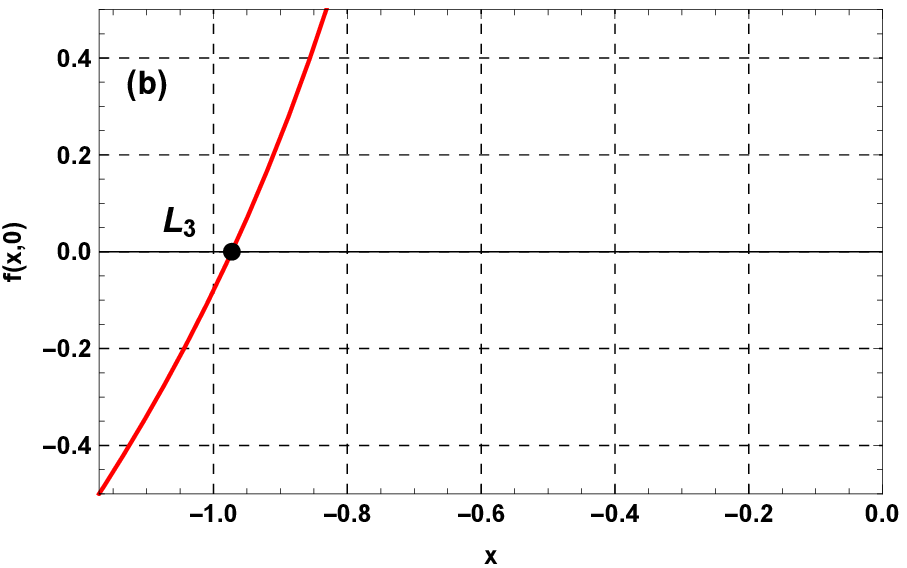} \hspace{2em}
	\caption{Collinear equilibrium point $L_{1}$, $L_{2}$ and $L_{3}$ in the Proxima Centauri system with dust belt.}\label{fig:onea}
\end{figure*}[h]
\begin{figure*}
	\includegraphics[width=0.45\textwidth]{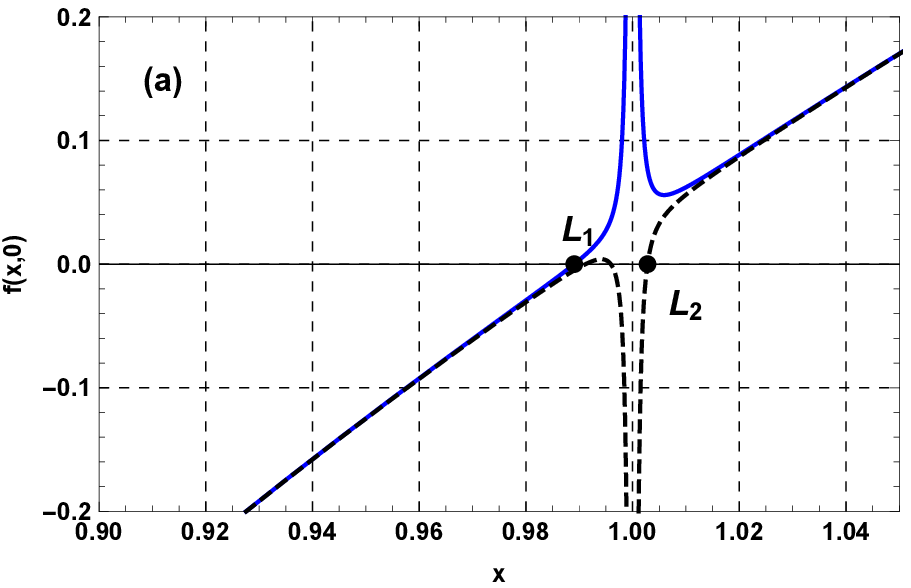} \hspace{2em}
	\includegraphics[width=0.45\textwidth]{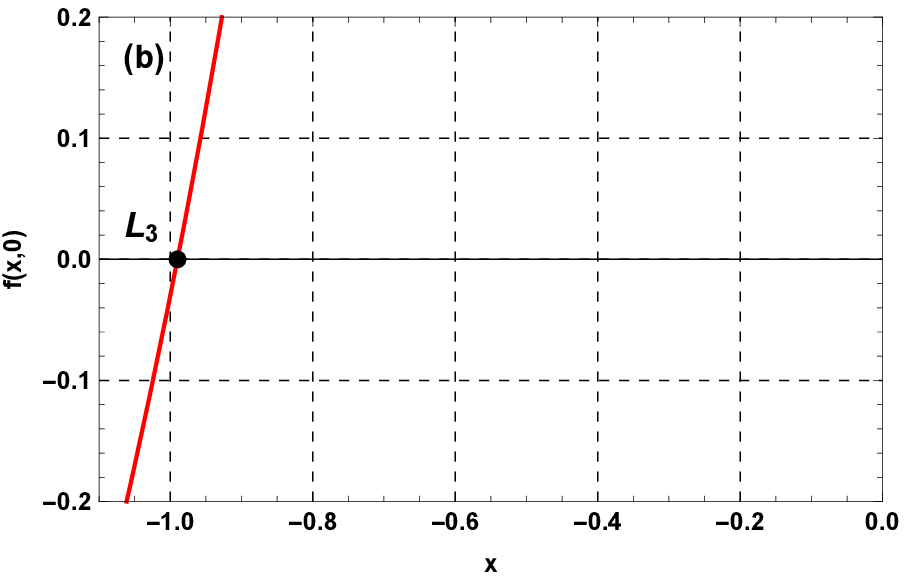} \hspace{2em}
	\caption{Collinear equilibrium point $L_{1}$, $L_{2}$ and $L_{3}$ in the Sun-Mars system with asteroid belt.}\label{fig:oneb}
\end{figure*}
\begin{figure*}[h]
	\includegraphics[width=0.45\textwidth]{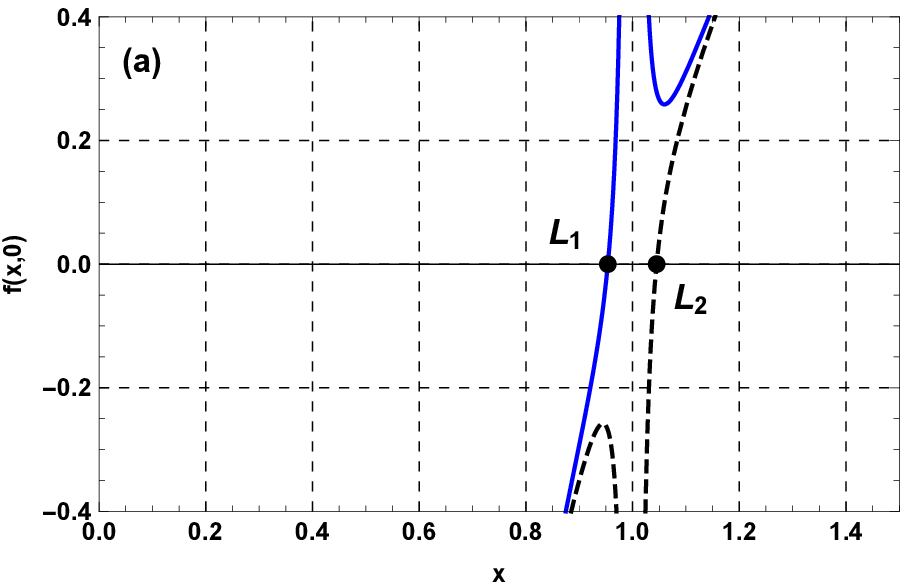} \hspace{2em}
	\includegraphics[width=0.45\textwidth]{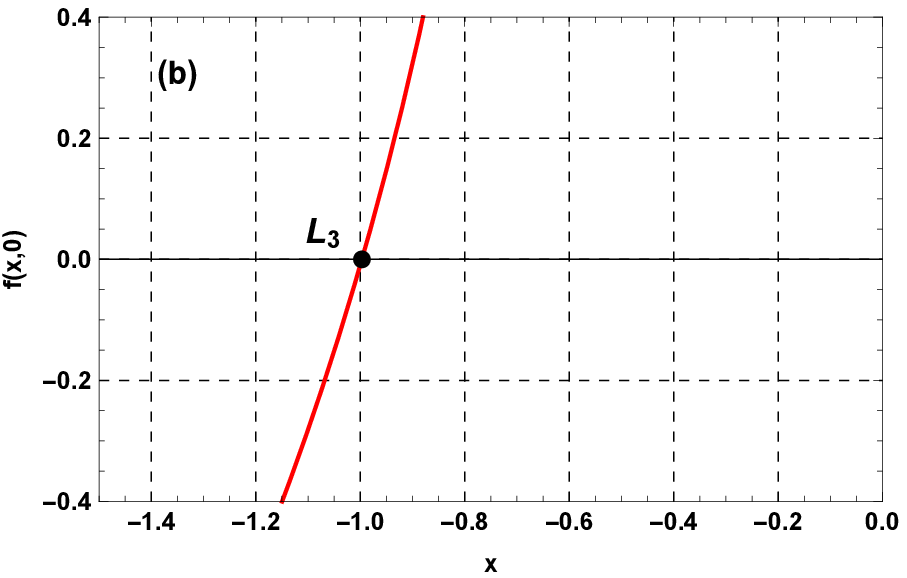} \hspace{2em}
	\caption{Collinear equilibrium point $L_{1}$, $L_{2}$ and $L_{3}$ in the Sun-Saturn system with Kuiper belt.}\label{fig:onec}
\end{figure*}
\begin{figure}
	\includegraphics[width=0.48\textwidth]{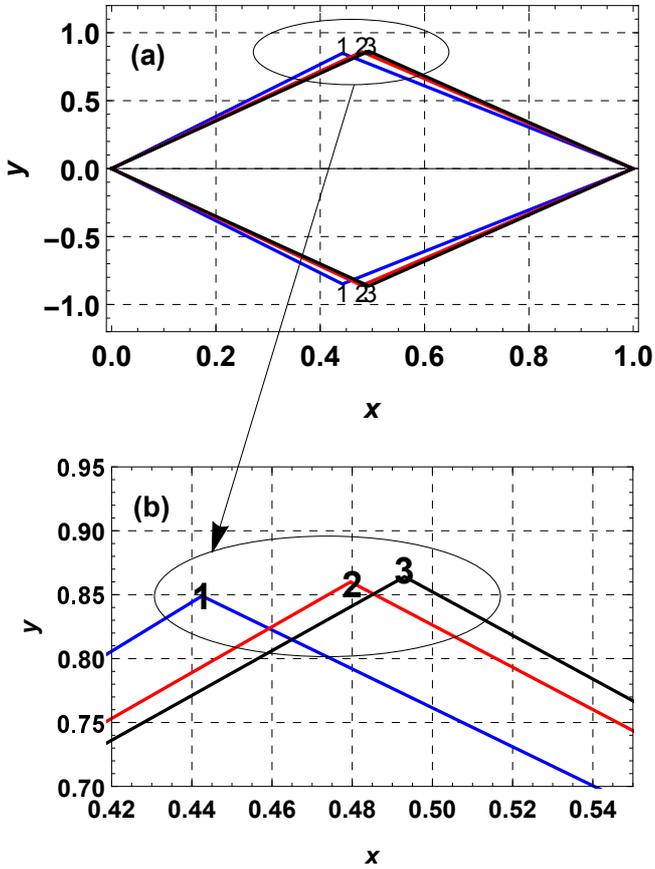}
	\caption{(a) Non-collinear equilibrium point $L_{4,5}$ in $(1)$ Proxima Centauri system with dust belt, $(2)$ Sun-Mars system with asteroid belt, $(3)$ Sun-Saturn system with Kuiper belt,  and  (b) Enlarge of oval-shaped region in part (a).}\label{fig:L45}
\end{figure}
Since, possible range of the locations of collinear equilibrium points are $(1-\mu,\,\infty)$, $(-\mu,\,1-\mu)$ and $(-\infty,\,-\mu)$ therefore, we have divided the collinear axis into three parts as $1-\mu<x<\infty$, $-\mu<x<1-\mu$ and $-\infty<x<-\mu$, consequently, equation (\ref{eq:cpt}) changes its form in the respective regions as    
\begin{eqnarray}
&&n^2x-\frac{q\left(1-\mu\right)}{(x+\mu)^2}\left\{1+\frac{3A_{21}}{2(x+\mu)^2}\right\}-\nonumber\\&& \frac{Q_{A}\mu}{(x+\mu-1)^2}\left\{1+\frac{3A_{22}}{2(x+\mu-1)^2}\right\}-\nonumber\\&&\frac{M_{d}x}{\left(x^2+T^2\right)^{\frac{3}{2}}}=0,\label{eq:cl1}\end{eqnarray}
\begin{eqnarray}
&&n^2x-\frac{q\left(1-\mu\right)}{(x+\mu)^2}\left\{1+\frac{3A_{21}}{2(x+\mu)^2}\right\}+\nonumber\\&&\frac{Q_{A}\mu}{(x+\mu-1)^2}\left\{1+\frac{3A_{22}}{2(x+\mu-1)^2}\right\}-\nonumber\\&&\frac{M_{d}x}{\left(x^2+T^2\right)^{\frac{3}{2}}}=0,\label{eq:cl2}\end{eqnarray}
\begin{eqnarray}
&&n^2x+\frac{q\left(1-\mu\right)}{(x+\mu)^2}\left\{1+\frac{3A_{21}}{2(x+\mu)^2}\right\}+\nonumber\\&&\frac{Q_{A}\mu}{(x+\mu-1)^2}\left\{1+\frac{3A_{22}}{2(x+\mu-1)^2}\right\}-\nonumber\\&&\frac{M_{d}x}{\left(x^2+T^2\right)^{\frac{3}{2}}}=0.\label{eq:cl3}\end{eqnarray} 
\begin{table*}
	\centering
		\caption{Coordinates of equilibrium points in different system.}\label{tab1}
		\begin{tabular}{@{}|lcccc|@{}} 
			\hline
			Planetary System &$L_{1}$ &$L_{2}$ & $L_{3}$ & $L_{4,5}$\\
			\hline
			Proxima Centauri with dust belt & 0.964443 & 1.015770 & -0.972602 & 0.442536,\,$\pm$ 0.848967\\\hline
			Sun-Mars with asteroid belt & 0.989063 & 1.002800 & -0.989898 & 0.479578,\,$\pm$ 0.859994\\\hline
			Sun-Saturn with Kuiper belt & 0.953561 & 1.045030 & -0.996775 & 0.493047,\,$\pm$ 0.864060\\\hline
		\end{tabular}
\end{table*}
Real solution of equations (\ref{eq:cl1}),\, (\ref{eq:cl2}) and (\ref{eq:cl3}) give the position of collinear equilibrium points $L_1,\,L_2$ and $L_3$, respectively. 
We have solved these three equations numerically for $L_1$, $L_2$ and $L_3$ at the approximate value of perturbation parameters, which are obtained from different source    \citep{bixel2017probabilistic,genio2017,anglada2017alma,nasafactsheet} and which are given below in the form of dimensionless quantity relative to different system as:
\begin{enumerate}
	\item[i.] Proxima Centauri system with dust disc: $\mu=0.000031,\, r_c=8,\, T=0.11,\, q=0.92,\, Q_{A}=0.9992,\, A_{21}=4.79\times10^{-6},\, A_{22}=2.21\times10^{-7}$ and $M_{d}=2.50\times10^{-7}$. 
	\item[ii.] Sun-Mars system with asteroid belt: $\mu=0.0000003,\, r_c=0.8,\, T=0.11,\, q=0.97, \,Q_{A}=0.9997,\, A_{21}=1.03\times10^{-9},\, A_{22}=5.21\times10^{-13}$ and $M_{d}=1.6\times10^{-9}$.
	\item[iii.] Sun-Saturn system with Kuiper belt: $\mu=0.000286,\, r_c=4.7,\, T=0.11,\, q=0.99, \,Q_{A}=0.9999, \,A_{21}=2.60\times10^{-11},\, A_{22}=6.59\times10^{-11}$ and $M_{d}=3.00\times10^{-7}$.
\end{enumerate}   
The resulting values are given in Table \ref{tab1}. Graphical solutions are also shown in Figure  \ref{fig:onea}-\ref{fig:onec}. In order to analyze the effect of perturbing parameters, we have obtained collinear as well as triangular equilibrium points at different values of $q$, $Q_A$, $A_{21}$, $A_{22}$ and $M_d$ for Proxima Centauri system with dust belt (Table \ref{tab2}-\ref{tab3}). From Table \ref{tab2}, it is noticed that oblateness of second primary have a significant effect on the positions of collinear equilibrium points $L_{1}$, $L_{2}$, and $L_{3}$ compare to that of first primary. The effect of radiation pressure and albedo on the position of $L_{1},\,L_{2}$ plays a significant role, whereas in case of $L_1$ effect of albedo is much less as compare to that of radiation pressure. A remarkable change in the position of $L_{1},\,L_{2},\,L_{3}$ due to the presence of disc in the problem is found. Form the results placed in the Table \ref{tab2}, it is clear that on increase in the value of $A_{21}$, $L_1$ moves away from the origin, $L_2$ shifts towards the origin, whereas position of $L_3$ remains unchanged. On the other hand, on increment in the value of $A_{22}$, $L_1$ and $L_2$ move towards and away from the origin, respectively, whereas no change in case of $L_3$. A decrease in $q$ results that a shift in the positions of $L_{1},\,L_{2},\,L_{3}$ towards the origin, whereas due to decrease in $Q_A$, $L_1$ shifts away from the origin, $L_2,\,L_3$ moves towards the origin. Due to increase in $M_d$, $L_{1}$ and $L_{2}$ move towards the origin, whereas $L_3$ shifts away from the origin. Thus , all the perturbing parameter have a considerable effect hence, these cannot be ignore during the mission design in space.

\subsection{Non-collinear equilibrium points $(L_{4,5})$}
\label{sec:triangular}

Since, for the non-collinear equilibrium points, $y\neq 0$. Therefore, by solving equations (\ref{eq:omx}) and (\ref{eq:omy}) with $y\neq 0$, we obtain the positions of non-collinear equilibrium points $L_{4,5}$. It is well known that in classical case, $r_{1}=1$ and $r_{2}=1$. So, in perturbed case, we assume that 
\begin{equation} 
r_{1}=1+\eta_{1} \quad \text{and} \quad r_{2}=1+\eta_{2}, \label{r1_r2}
\end{equation}
where $\eta_{1}$ and $\eta_{2}$ are very small real quantity. Taking only first order terms of $\eta_{1}$ and $\eta_{2}$ in the Taylor's series expansion of the expressions of $r_1$ and $r_2$ together with equations (\ref{eq:omx}) and (\ref{eq:omy}), we have obtained the position of non-collinear equilibrium points as
\begin{eqnarray}
x&=&\frac{1}{2}-\mu+2\left(\eta_{1}-\eta_{2}\right),\label{eq:x}
\end{eqnarray}
\begin{eqnarray}
y&=&\pm \frac{1}{2}\sqrt{3+4\left(\eta_{1}+\eta_{2}\right)},\label{eq:y}
\end{eqnarray}
where
\begin{eqnarray*}
	&&\eta_{1}=\frac{\left[ n^2-q\left(1+\frac{3A_{21}}{2}\right)-\frac{M_{d}}{(r_{c}^2+T^2)^{\frac{3}{2}}}\right]}{-3q(1+\frac{5A_{21}}{2})},\\&&
	\eta_{2}=\frac{\left[n^2-Q_{A}\left(1+\frac{3A_{22}}{2}\right)-\frac{M_{d}}{(r_{c}^2+T^2)^{\frac{3}{2}}}\right]}{{-3Q_{A}(1+\frac{5A_{22}}{2})}}.
\end{eqnarray*}

\begin{table*}
	\centering
	\caption{Coordinates of $L_{1,2,3}$ at $\mu=0.000031$, $T=0.11$ for different values of $q$, $Q_A$, $A_{21}$, $A_{22}$ and $M_d$.}\label{tab2}
	\begin{tabular}{@{}|cccccccc|@{}} 
		\hline
		$q$ & $Q_{A}$ & $A_{21}$ & $A_{22}$ & $M_{d}$ & $L_1:(x,\,0)$ & $L_2:(x,\,0)$ & $L_3:(x,\,0)$ \\	
		\hline
		1.00 & 1.0000 & 0 & 0 & 0 & 0.978347009 & 1.02190724 & -1.0000129\\
		0.99 & 1.0000 & 0 & 0 & 0 & 0.977157185 & 1.02086380 & -0.9966684\\
		0.98 & 1.0000 & 0 & 0 & 0 & 0.975832539 & 1.01991698 & -0.9933013\\
		0.98 & 0.9996 & 0 & 0 & 0 & 0.975835371 & 1.01991405 & -0.9933013\\
		1.00 & 1.0000 & $4.8\times10^{-6}$ & 0 & 0 & 0.978347097 & 1.02190716 & -1.0000129\\ 
		1.00 & 1.0000 & 0 & $2.21\times10^{-7}$ & 0 & 0.978341905 & 1.02191227 & -1.0000128\\  
		1.00 & 1.0000 & 0 & 0 & $2.5\times10^{-7}$ & 0.978346982 & 1.02190721 & -1.0000131\\
		0.98 & 0.9996 &  $4.8\times10^{-6}$ &$2.21\times10^{-7}$ & $2.5\times10^{-7}$ & 0.975831369 & 1.01992001 & -0.9933015\\\hline
	\end{tabular}
\end{table*}

In the absence of all perturbations, the positions of $L_{4,5}$ as in equations (\ref{eq:x}) and (\ref{eq:y}), agree with that of classical values as $x=\frac{1}{2}-\mu$ and $y=\pm \frac{\sqrt{3}}{2}$. Also, we have computed the non-collinear equilibrium points numerically at different value of parameters $q,\,Q_A,\, A_{21},\,A_{22}$ and $M_d$ for different planetary systems (Table \ref{tab1}). Moreover, to analyze the effect of perturbing parameters on $L_{4,5}$, we have obtained the value of $L_{4,5}$ for Proxima Centauri system with dust belt at different value of perturbing parameters (Table  \ref{tab3}). From Table \ref{tab3}, it is observed that the effect of oblateness of the primaries is less on the positions of non-collinear equilibrium points $L_{4,5}$, whereas effect of radiation pressure and albedo are significant. A considerable effect of the disc can also be observed. Form Table \ref{tab3}, it is noticed that on increase in the value of $A_{21}$ and $A_{22}$, $y$ coordinates of $L_{4,5}$ shift slightly towards the origin, whereas $x$ coordinates move away from and towards the origin, respectively. Alike as in case of oblateness of the primaries, on decrement in the values of $q$ and $Q_A$, $y$ coordinates move towards the origin, whereas $x$ coordinates shift towards the origin in either case and away from the origin in later case. However, on increase in the value of $M_d$, $x$ coordinates as well as $y$ coordinates move towards the origin. Thus, the effect of all perturbing parameter are not ignorable during the study of motion in the neighborhood of $L_{4,5}$.

\begin{table*}
	\centering
	\caption{Coordinates of $L_{4,5}:(x_{4,5},\,\pm y_{4,5})$ at $\mu=0.000031$ ,$r_{c}=8$, $T=0.11$ for different values of $q$, $Q_A$, $A_{21}$, $A_{22}$ and $M_d$.}\label{tab3}
	\begin{tabular}{@{}|ccccccc|@{}} 
		\hline
		$q$ & $Q_{A}$ & $A_{21}$ & $A_{22}$ & $M_{d}$ & $x_{4,5}$ & $\pm y_{4,5}$ \\	
		\hline
		1.00 & 1.0000 & 0 & 0 & 0 & 0.4999690 & 0.8660254 \\
		0.99 & 1.0000 & 0 & 0 & 0 & 0.4966301 & 0.8640891 \\
		0.98 & 1.0000 & 0 & 0 & 0 & 0.4932799 & 0.8621291 \\
		0.98 & 0.9996 & 0 & 0 & 0 & 0.4934132 & 0.8620525 \\
		1.00 & 1.0000 & $4.8\times10^{-6}$ & 0 & 0 &  0.4999714 & 0.8660240 \\ 
		1.00 & 1.0000 & 0 & $2.21\times10^{-7}$ & 0 & 0.4999688 & 0.8660253 \\  
		1.00 & 1.0000 & 0 & 0 & $2.5\times10^{-7}$ &  0.4999689 & 0.8660253 \\
		0.98 & 0.9996 &  $4.8\times10^{-6}$ &$2.21\times10^{-7}$ & $2.5\times10^{-7}$ & 0.4934156 & 0.8620511 \\\hline
	\end{tabular}
\end{table*}

\section{Analysis of linear stability}
\label{sec:linearstab}

Since, it is known that if we displace a test particle from its initial point by giving a very small velocity then either particle will oscillate around the initial point or it will depart from that point. If the particle oscillate around the initial point for a considerable period of time then we say that the initial point is stable otherwise it is said to be unstable \citep{murray1999,moulton2012introduction}. In order to examine the linear stability in the vicinity of the equilibrium points, we linearized the equations of motion of infinitesimal mass around initial point. Suppose, $(x_{e},y_{e})$ is the coordinate of the initial point, which is one of the equilibrium points $L_i,\,i=1, 2, 3, 4, 5$ of the RTBP. Let $X=Ae^{\lambda t}$ and $Y=Be^{\lambda t}$ be the small displacements of the infinitesimal mass from the point $(x_{e},\,y_{e})$, where $\lambda$ is a parameter and $A,\,B$ are constants to be determined, then the final position of the infinitesimal mass will be $x=x_{e}+X$ and $y=y_{e}+Y$. Substituting these values of $x$ and $y$ in equation (\ref{eq:ox}) and (\ref{eq:oy}), we get
\begin{equation}
\ddot{X}-2n\dot{Y}=\Omega_{x}(x_{e}+X,y_{e}+Y),\label{eq:1}
\end{equation} 
\begin{equation}
\ddot{Y}+2n\dot{X}=\Omega_{y}(x_{e}+X,y_{e}+Y).\label{eq:2}
\end{equation}	
Expanding the right hand sides of the equations (\ref{eq:1}) and (\ref{eq:2}) about $(x_{e},y_{e})$ using Taylor's series expansion and then considering only first order terms in $X$ and $Y$, we find the linearized equations of motion of the infinitesimal mass in the neighborhood of equilibrium point as
\begin{equation}
\ddot{X}-2n\dot{Y}=X\Omega_{xx}^e+Y\Omega_{xy}^e,\label{eq:X}
\end{equation} 
\begin{equation}
\ddot{Y}+2n\dot{X}=X\Omega_{yx}^e+Y\Omega_{yy}^e,\label{eq:Y}
\end{equation}
where $\Omega_{xx}^e$, $\Omega_{xy}^e$, $\Omega_{yx}^e$ and $\Omega_{yy}^e$ are second-order partial derivatives of $\Omega$ at equilibrium point $(x_{e},y_{e})$, which are obtained form equation (\ref{eq:ep}). Equations (\ref{eq:X}) and (\ref{eq:Y}) show that the net force on infinitesimal mass in the neighborhood of the point $(x_{e},\,y_{e})$, is directly proportional to the displacement.
Now, using $X=Ae^{\lambda t}$ and $Y=Be^{\lambda t}$ in the equations (\ref{eq:X}) and (\ref{eq:Y}), we get
\begin{eqnarray}
(\lambda^2-\Omega_{xx}^e)A+(-2n\lambda-\Omega_{xy}^e)B&=&0,\label{eq:linearsys1}\\
(2n\lambda-\Omega_{yx}^e)A+(\lambda^2-\Omega_{yy}^e)B&=&0.\label{eq:linearsys2}
\end{eqnarray}
The above system of linear equations has non-trivial solution if
\begin{equation}
\begin{vmatrix} 
\lambda^2-\Omega_{xx}^e & -2n\lambda-\Omega_{xy}^e \\
2n\lambda-\Omega_{yx}^e & \lambda^2-\Omega_{yy}^e 
\end{vmatrix}=0.\label{eq:deter}
\end{equation}
Simplifying above determinant, we get a bi-quadratic equation in $\lambda$ as
\begin{equation} 
\lambda^4+C\lambda^2+D=0,\label{eq:char}
\end{equation}where 
\begin{equation}
C=4n^2-(\Omega_{xx}^e+\Omega_{yy}^e) \quad \text{and} \quad D=\Omega_{xx}^e\Omega_{yy}^e-(\Omega_{xy}^e)^2.\label{eq:cd}
\end{equation}
Equation (\ref{eq:char}) is known as characteristic equation of the system.
Let $\lambda_{1}$, $\lambda_{2}$, $\lambda_{3}$  and $\lambda_{4}$ be the four roots of the characteristic equation (\ref{eq:char}), which are given as
\begin{equation}
\lambda_{1,2,3,4}=\pm\left(\frac{-C\pm \sqrt{C^2-4D}}{2}\right)^{\frac{1}{2}},\label{eq:roots}
\end{equation} 
then general solution of the system of linear differential equations (\ref{eq:X}) and (\ref{eq:Y}) with constant coefficients can be written as:
\begin{eqnarray}
X(t)=A_{1}e^{\lambda_{1}t}+A_{2}e^{\lambda_{2}t}+A_{3}e^{\lambda_{3}t}+A_{4}e^{\lambda_{4}t},\label{eq:solx}\\
Y(t)=B_{1}e^{\lambda_{1}t}+B_{2}e^{\lambda_{2}t}+B_{3}e^{\lambda_{3}t}+B_{4}e^{\lambda_{4}t},\label{eq:soly}
\end{eqnarray}
where constants $B_{1}$, $B_{2}$, $B_{3}$, $B_{4}$ are related to four arbitrary constants $A_{1}$, $A_{2}$, $A_{3}$, $A_{4}$, respectively, by the means of equations (\ref{eq:linearsys1}) and (\ref{eq:linearsys2}). From the equations (\ref{eq:solx}) and (\ref{eq:soly}), it is clear that if the four roots $\lambda_{j},j=1,2,3,4$ are purely imaginary then the solution $X$ and $Y$ can be written in the form of periodic function, consequently, it will be stable. If all the roots are multiple purely imaginary, then the existence of secular term in the solution makes it unstable. when all the roots are either real or complex with at least one positive real part then solution contains exponential term which makes  it unstable. If the real parts of all roots are negative then solution will be asymptotically stable \citep{boccaletti1996integrable,kishor2013linear}.

\subsection{Linear Stability of collinear equilibrium points}
\label{subsec:stbcoll}

\begin{figure}
	\centering
	{\includegraphics[width = .45\textwidth]{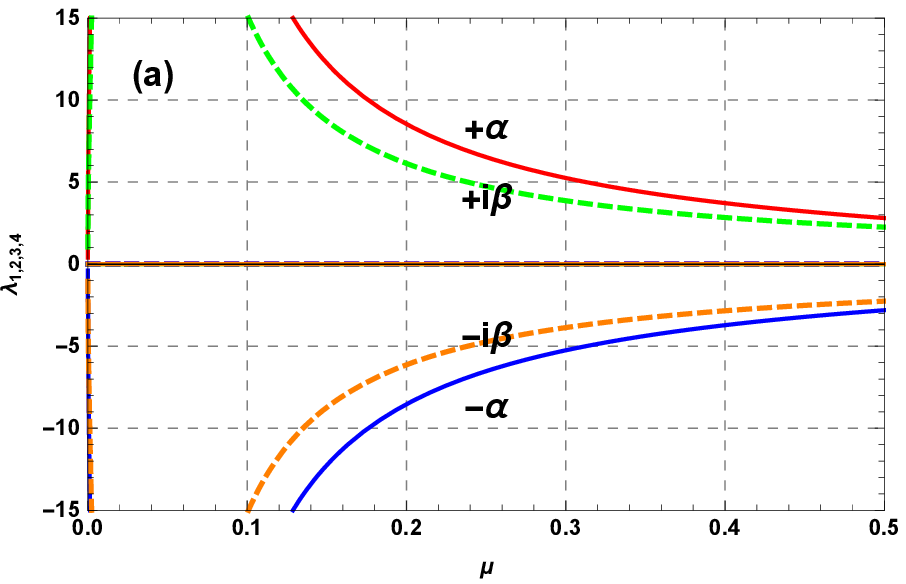}}\\\vspace{2em}
	{\includegraphics[width = .45\textwidth]{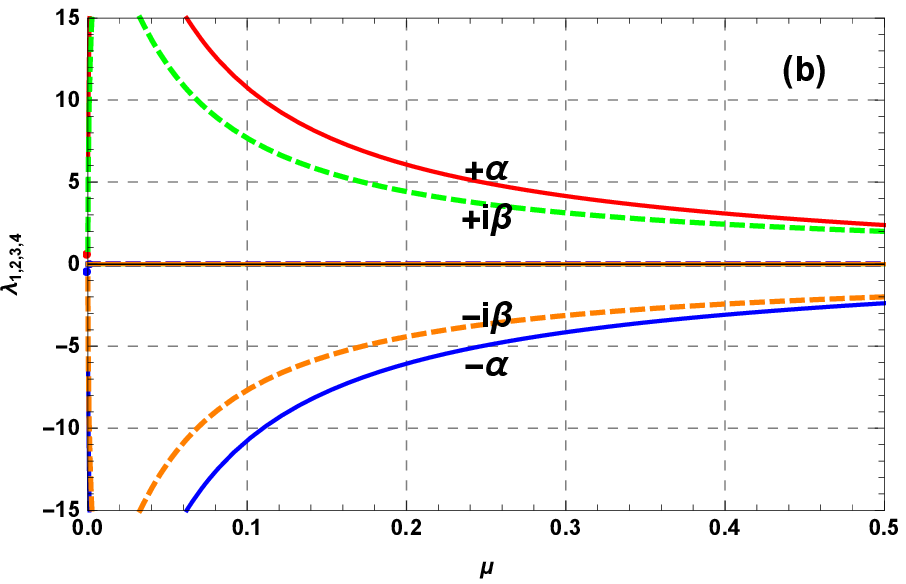}}\\\vspace{2em}
	{\includegraphics[width = .45\textwidth]{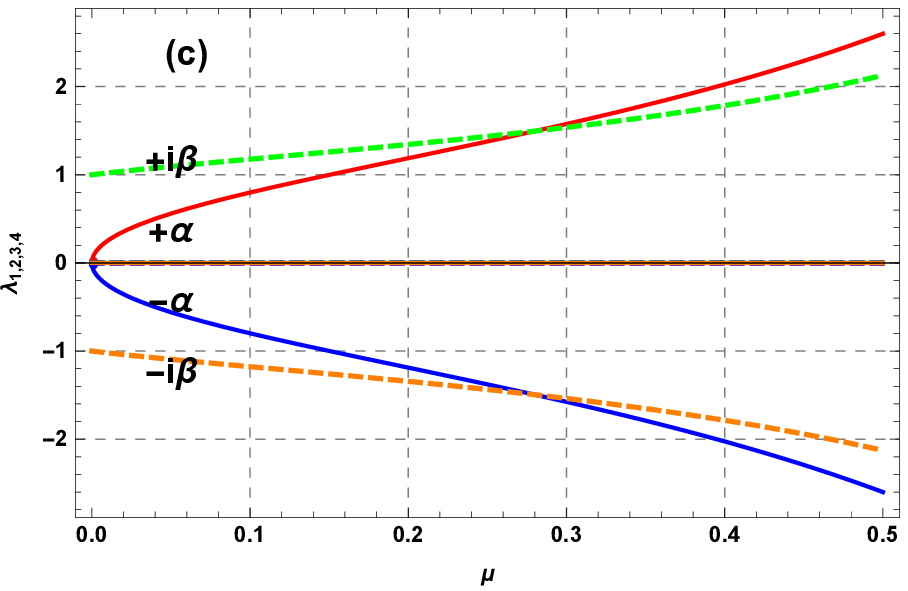}} 
	\caption{Variation of characteristic roots for collinear equilibrium points  (a) $L_{1},\,$ (b) $L_2,\,$ (c) $L_3$ with respect to mass parameter $(\mu)$ in Proxima Centauri system.}
	\label{fig:mL123}
\end{figure}

As, it is known that generally all the collinear equilibrium points of the RTBP are unstable, yet we have examined the linear stability of $L_{1},\,L_{2},\,L_{3}$ with respect to parameter $\mu$ and $Q_A$. Within the range of mass parameter, $0< \mu < 0.5$, we found that $\Omega_{xx}^e>0$ and $\Omega_{yy}^e<0$ for all the collinear equilibrium points $L_{1},\,L_{2},\,L_{3}$. Consequently, the discriminant $C^2-4D>0$ and hence, the characteristic equation (\ref{eq:char}) gives at least one positive real root. Due to one positive real root (see Figure \ref{fig:mL123}) for each collinear equilibrium points, solutions (\ref{eq:solx}-\ref{eq:soly}) become unstable, which insure the instability of all the collinear equilibrium points. We also, have analyzed the linear stability relative to perturbation parameter $q$,  $Q_A$, $A_{21}$, $A_{22}$ and $M_{d}$ on similar basis in the possible range and found that there no effect on the instability of each collinear equilibrium points and hence, these are again unstable.

\subsection{Linear stability of non-collinear equilibrium points} 
\label{subsec:stbnoncoll}

In this analysis linear stability of $L_4$ is discussed, whereas dynamics of $L_5$ is quite similar to that of $L_4$. As, we know that to insure the linear stability of $L_4$, motion of infinitesimal mass about it must be bounded and periodic. In other words,  all four roots of the characteristic equation (\ref{eq:char}) must be purely imaginary and this can be examined by observing the sign of the discriminant. Therefore the sign of the discriminant $(C^2-4D)$ of the characteristic equation (\ref{eq:char}) defines the nature of roots. For stable motion, discriminant $C^2-4D$ must be positive and $C^2>4D>0$ so that all four roots will be purely imaginary. Moreover, the solution of $C^2-4D=0$ gives the value of  critical mass ratio, consequently, the range of stability and instability. That is, there are three cases (i) $C^2-4D<0$, (ii) $C^2-4D=0$, and (iii) $C^2-4D>0$ to describe the linear stability of non-collinear equilibrium point, which correspond to the three ranges (i) $\mu_c<\mu<0.5$, (ii) $\mu=\mu_c$, and (iii) $0<\mu<\mu_c$, respectively, of the mass parameter in terms of critical mass ratio $\mu_c$. Thus, before going in detail, we compute the critical mass ratio $\mu_c$.
For this, we solve the discriminant $C^2-4D$ of the characteristic equation (\ref{eq:char}) by equating zero. That is
\begin{equation}
(4n^2-\Omega_{xx}^e-\Omega_{yy}^e)^2-4\left\{\Omega_{xx}^e\Omega_{yy}^e-(\Omega_{xy}^e)^2\right\}=0.\label{eq:discri}
\end{equation}
As, $0<q$, $Q_A\leq 1$, so to minimize the complexity, we take $q=1-\epsilon_{1}$ and $Q_{A}=1-\epsilon_{2}$, where $0\leq\epsilon_{1},\,\epsilon_{2}<1$. Using Taylor's series expansion in (\ref{eq:discri}), we obtain critical mass ratio $(\mu_{c})$ in terms of all perturbing parameters as
\begin{eqnarray}
\mu_{c}&=&0.0385209+0.2163963\,\epsilon_{1}-0.1539740\,\epsilon_{2}+\nonumber\\&&0.2948124\,A_{21}+0.6281458\, A_{22}+\nonumber\\&&0.8877527\,M_{d}.\label{eq:cmr}
\end{eqnarray}In the expression (\ref{eq:cmr}), second and higher order terms have been neglected due to their very small contribution. Being the function of $q,\, Q_A,\, A_{21},\,A_{22}$ and $M_d$, expression (\ref{eq:cmr}), also called as perturbed mass ratio. In the absence of perturbations, expression (\ref{eq:cmr}) agree with that of classical value $\mu_{c}=0.0385209$ \citep{Szebehely1967torp.book.....S,Deprit1967AJ.....72..173D}. In the presence of perturbations, we have computed critical mass ratio ($\mu_c$)  for Proxima Centauri system with dust belt at different values of $q,\, Q_A,\, A_{21},\,A_{22}$ and $M_d$ (see Table \ref{tab4}). From Table \ref{tab4}, it is observed that critical mass ratio $\mu_c$ increases with the increase in the values of $\epsilon_1=1-q,\,A_{21},\,A_{22}$ and $M_d$, however it decreases with the increment in the value of $\epsilon_2=1-Q_A$. The detail analysis under three cases for $L_4$ in Proxima Centauri system with dust belt are as follows, whereas analysis of $L_5$ is similar to that of $L_4$.

\begin{table*}
	\centering
	\small
	\caption{Critical mass ratio $\mu_{c}$ for Proxima Centauri system at different values of $q$, $Q_A$, $A_{21}$, $A_{22}$ and $M_d$.}\label{tab4}
	\begin{tabular}{@{}|cccccc|@{}} 
		\hline
		$\epsilon_1$ & $\epsilon_2$ & $A_{21}$ & $A_{22}$ & $M_{d}$ & $\mu_c$  \\	
		\hline
		0 & 0 & 0 & 0 & 0 & 0.0385209 \\
		0.01 & 0 & 0 & 0 & 0 & 0.0406849  \\
		0.02 & 0 & 0 & 0 & 0 & 0.0428488  \\
		0.02 & 0.0002 & 0 & 0 & 0 & 0.0428167  \\
		0.02 & 0.0004 & 0 & 0 & 0 & 0.0427845  \\
		0 & 0 & $2.8\times10^{-6}$ & 0 & 0 &  0.0385217  \\
		0 & 0 & $4.8\times10^{-6}$ & 0 & 0 &  0.0385223  \\ 
		0 & 0 & 0 & $1.2\times10^{-7}$ & 0 & 0.0385210  \\
		0 & 0 & 0 & $2.2\times10^{-7}$ & 0 & 0.0385211  \\
		0 & 0 & 0 & 0 & $1.5\times10^{-7}$ &  0.0385209  \\
		0 & 0 & 0 & 0 & $2.5\times10^{-7}$ &  0.0385208  \\
		0.02 & 0.0004 &  $4.8\times10^{-6}$ &$2.21\times10^{-7}$ & $2.5\times10^{-7}$ & 0.0427863  \\\hline
	\end{tabular}
\end{table*}

\textbf{Case I:} When $C^2-4D<0$ i.e. $\mu_{c}<\mu\leq 0.5$:\\
Due to negative discriminant, roots $\lambda_{1,2,3,4}$ defined in equation (\ref{eq:roots}) reduces to complex form as
\begin{eqnarray*}
	\lambda_{1}&=\frac{1}{\sqrt{2}}(-C+ i\Delta)^\frac{1}{2}=a_{1}+ib_{1},\\
	\lambda_{2}&=-\frac{1}{\sqrt{2}}(-C+ i\Delta)^\frac{1}{2}=a_{2}+ib_{2},\\
	\lambda_{3}&=\frac{1}{\sqrt{2}}(-C- i\Delta)^\frac{1}{2}=a_{3}+ib_{3},\\
	\lambda_{4}&=-\frac{1}{\sqrt{2}}(-C- i\Delta)^\frac{1}{2}=a_{4}+ib_{4}.\\
\end{eqnarray*}
with $\Delta=\sqrt{C^2-4D}$. Due to the presence of positive real parts in two roots,  motion of the infinitesimal mass in the neighborhood of $L_4$ becomes unstable. Hence, $L_4$ is unstable for $\mu_{c}<\mu\leq 0.5$.

\textbf{Case II:} When $C^2-4D=0$ i.e. $\mu=\mu_{c}$:

In this case, roots (\ref{eq:roots}) of the characteristic equation (\ref{eq:char}) take the form
\begin{equation}
\lambda_{1,3}=\pm i\sqrt{\frac{C}{2}},\quad \lambda_{2,4}=\pm i\sqrt{\frac{C}{2}},
\end{equation}
which are multiple purely imaginary roots with equal magnitude and hence, existence of secular term in the solution of the equations of motion of infinitesimal mass in the neighborhood of $L_4$ is guaranteed, which insure the instability of equilibrium point $L_4$.

\begin{figure}
	\centering
	\includegraphics[width =0.47 \textwidth]{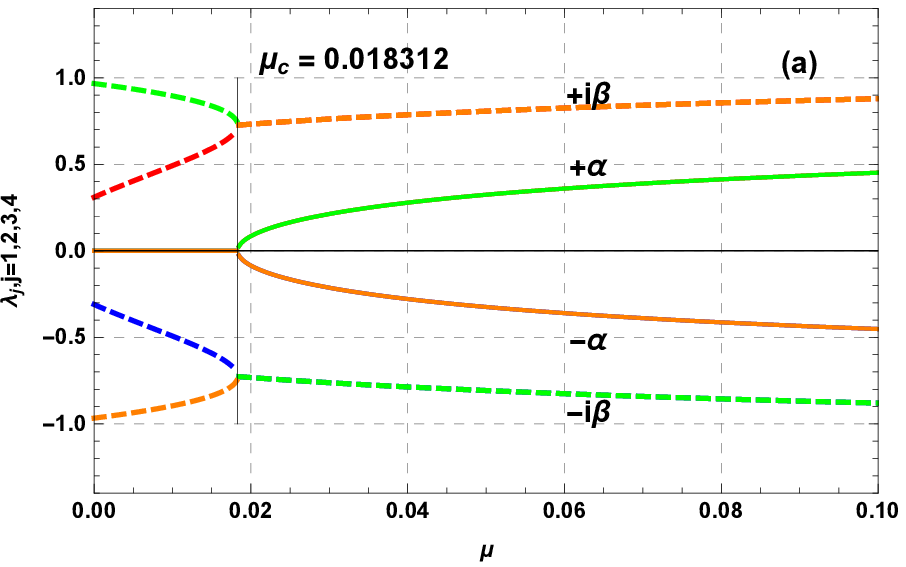}\\\vspace{2em}
	\includegraphics[width =0.47 \textwidth]{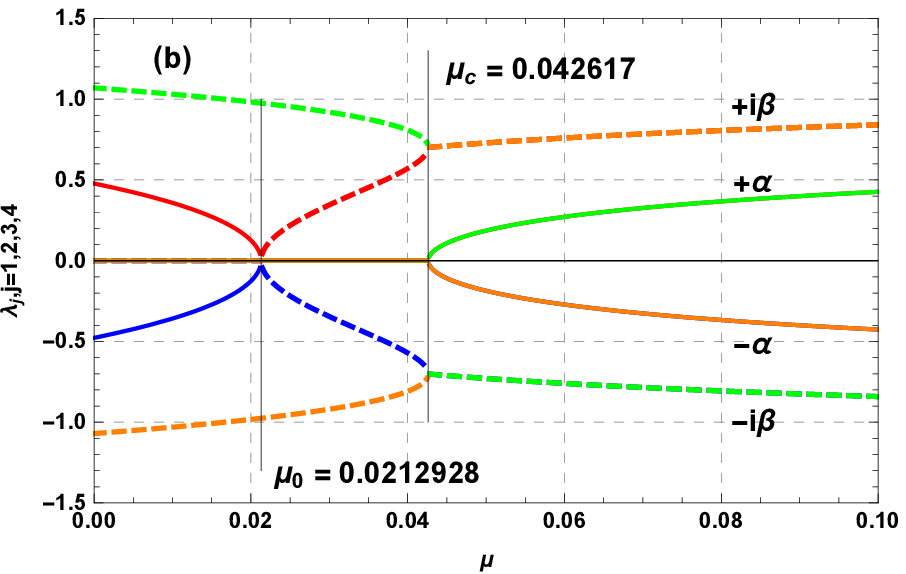}
	\caption{Real (bold line) and imaginary (dashed line) parts of the roots $\lambda_{1,2,3,4}$ for perturbed non-collinear equilibrium point with respect to mass ratio $\mu$ at: \textbf{(a)} $L_4\,(0.4934156,\,0.8620511)$ and \textbf{(b)} $L_4\,(0.442536,\,0.848967)$ in Proxima Centauri system with dust belt.}\label{rootL4a}
\end{figure}

\begin{figure}
	\centering
	\includegraphics[width =0.47 \textwidth]{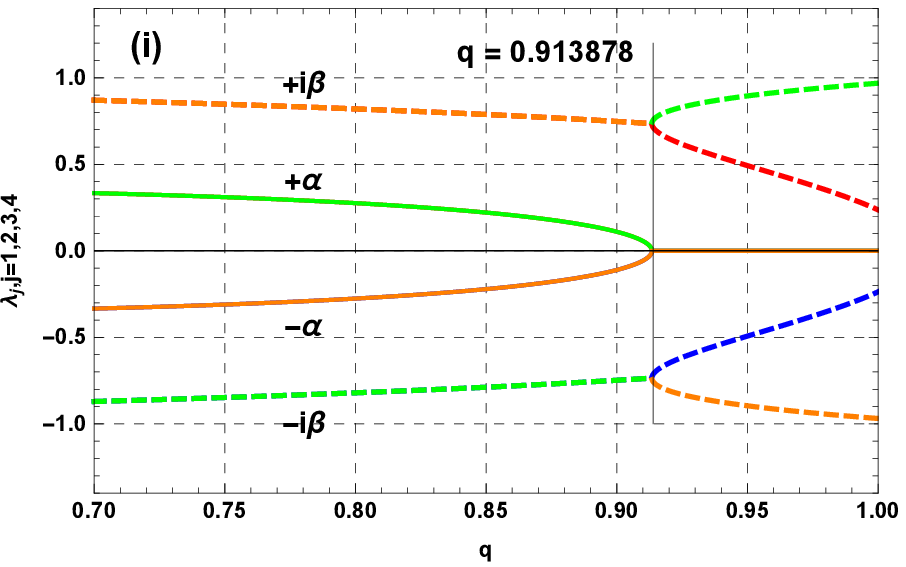}\\\vspace{2em}
	\includegraphics[width =0.47 \textwidth]{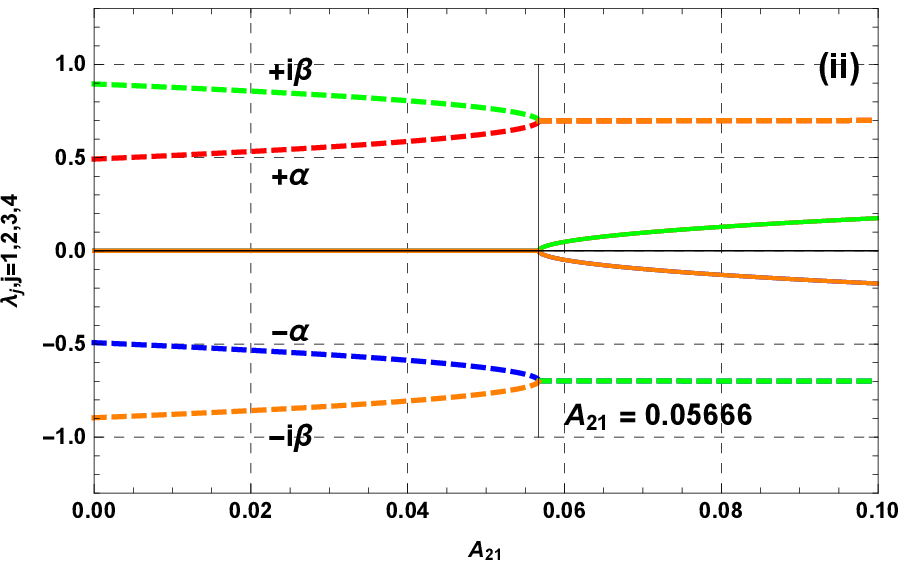}\\\vspace{2em}
	\includegraphics[width =0.47 \textwidth]{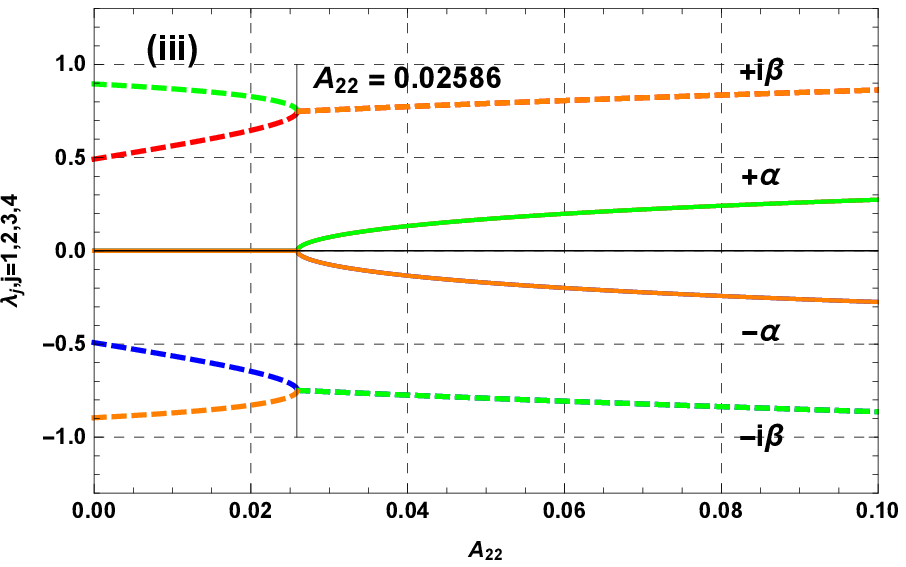}
	\caption{Real (bold line) and imaginary (dashed line) parts of the roots $\lambda_{1,2,3,4}$ for perturbed non-collinear equilibrium point with respect to: \textbf{(i)} $q$ at $\mu=0.01$, \textbf{(ii)} $A_{21}$ at $\mu=0.01$ and \textbf{(iii)} $A_{22}$ at $\mu=0.01$ in Proxima Centauri system with dust belt.}\label{rootL4b}
\end{figure}

\textbf{Case III:} When $C^2-4D>0$ i.e. $0<\mu<\mu_{c}$:

As, to insure the linear stability, four roots must be purely imaginary, which means that $C^2-4D>0$ and $C^2>4D>0$. We have analyzed this case for the Proxima Centauri system with dust belt and for particular  $L_4\,(\,0.4934156,\,0.8620511)$, we found that the above condition is true only for $0<\mu<\mu_{c}=0.0183124$. Further, if we take  $L_4\,(\,0.442536,\,0.848967)$, the $L_4$ becomes stable for $0.0212928=\mu_{0}<\mu<\mu_{c}=0.042617$ as shown in Fig. \ref{rootL4a}. Hence, motion of the infinitesimal mass is periodic, consequently $L_4\,(\,0.4934156,\,0.8620511)$ is stable, for $0<\mu<\mu_c$, whereas  $L_4\,(\,0.442536,\,0.848967)$ is stable only for $\mu_0<\mu<\mu_0$. Also, we have analyzed the effect of perturbation parameters $q$,\,$Q_{A}$,\,$A_{21}$,\,$A_{22}$,\,$M_{d}$ in the context of the nature of roots as shown in Fig. \ref{rootL4b}. A variation of linear stability range with respect to these parameters can be observed in the Fig. \ref{rootL4b}.

Now, for $0.0212928=\mu_0<\mu<\mu_{c}=0.042617$, four roots can be written as
\begin{eqnarray}
\lambda_{1,2}=\pm\left(\frac{-C+\sqrt{C^2-4D}}{2}\right)^{\frac{1}{2}}=\pm i\,\omega_{1},\label{eq:nrt1} 
\end{eqnarray}
\begin{eqnarray}
\lambda_{3,4}=\pm\left(\frac{-C- \sqrt{C^2-4D}}{2}\right)^{\frac{1}{2}}=\pm i\,\omega_{2},\label{eq:nrt2}
\end{eqnarray} where $i=\sqrt{-1}$.
Therefore, form equations (\ref{eq:solx}-\ref{eq:soly}) and (\ref{eq:nrt1}-\ref{eq:nrt2}), there are two periodic motions of the infinitesimal mass in the neighborhood of $L_4$ namely, short periodic and long periodic motions with periods of $\dfrac{2\pi}{\omega_{1}}$ and $\dfrac{2\pi}{\omega_{2}}$, respectively. The short periodic motion is found very nearer to the orbital period of the second primary,  whereas long periodic motion is related to the liberation in the neighborhood of $L_{4}$ \citep{murray1999}. Suppose, coefficients in equations (\ref{eq:solx}-\ref{eq:soly}) are  of the complex form like  $A_{j}=c_{j}+id_{j}$ and $B_{j}=e_{j}+if_{j},\, j=1,2,3,4$ such as  $c_{j},\,d_{j},\,e_{j},\,f_{j}\in \mathbb{R}$, then the solution $X(t)$ and $Y(t)$ becomes
\begin{eqnarray}
X(t)&=&(c_{1}+id_{1})e^{\lambda_{1}t}+(c_{2}+id_{2})e^{\lambda_{2}t}+\nonumber\\&&(c_{3}+id_{3})e^{\lambda_{3}t}+(c_{4}+iq_{4})d^{\lambda_{4}t},\label{eq:Xt}\end{eqnarray}
\begin{eqnarray}
Y(t)&=&(e_{1}+if_{1})e^{\lambda_{1}t}+(e_{2}+if_{2})e^{\lambda_{2}t}+\nonumber\\&&(e_{3}+if_{3})e^{\lambda_{3}t}+(e_{4}+if_{4})e^{\lambda_{4}t}.\label{eq:Yt}
\end{eqnarray}
As, the coefficients of the exponential, in equation (\ref{eq:Xt}) and (\ref{eq:Yt}) are in complex conjugate pairs, hence  we can take
\begin{eqnarray*}
	&&c_{1}=c_{2}=\alpha_{1}, c_{3}=c_{4}=\alpha_{2}, d_{1}=-d_{2}=\beta_{1}, \\&& d_{3}=-d_{4}=\beta_{2}
	e_{1}=e_{2}=\gamma_{1},  e_{3}=e_{4}=\gamma_{2}, \\&& f_{1}=-f_{2}=\delta_{1}, f_{3}=-f_{4}=\delta_{2} 
\end{eqnarray*}
Again, using Euler's relation $e^{i\theta}=\cos{\theta}+i\sin{\theta}$, solution $X(t)$ and $Y(t)$ of the infinitesimal mass reduces to real periodic solution as 
\begin{eqnarray}
X(t)&=&2\alpha_{1}\cos{\omega_{1}t}+2\alpha_{2}\cos{\omega_{2}t}-\nonumber\\&&2\beta_{1}\sin{\omega_{1}t}-2\beta_{2}\sin{\omega_{2}t},\label{eq:Xt1}\\
Y(t)&=&2\gamma_{1}\cos{\omega_{1}t}+2\gamma_{2}\cos{\omega_{2}t}-\nonumber\\&&2\delta_{1}\sin{\omega_{1}t}-2\delta_{2}\sin{\omega_{2}t},\label{eq:Yt1}
\end{eqnarray}
which shows that the motion of the infinitesimal mass in the vicinity of non-collinear equilibrium point are oscillatory hence, stable. As a specific example, consider the stability of the $L_{4}: (0.442536,\,0.848967)$ point for  $q=0.92$, $Q_{A}=0.9992$, $A_{21}=4.8\times10^{-6}$, $A_{22}=2.21\times10^{-7}$, $M_{d}=2.5\times10^{-7}$, $\mu_0<\mu=0.03<\mu_{c}$, $r_{c}=8$ and $T=0.11$. Assume that the small displacements are $X_0=Y_0=10^{-5}$ and initial velocity $\dot{X_0}=\dot{Y_0}=0$. We have found the resulting eigen values as $\lambda_{1,2}=\pm \omega_1 i=\pm 0.473\,i$ and $\lambda_{3,4}=\pm \omega_2 i=\pm 0.875\,i$,  respectively, which indicate that this point is stable due to pure imaginary eigen values. After computing $A_j$ and $B_j$ by the means of the equation (\ref{eq:linearsys1}), the orbits of infinitesimal mass in the vicinity of $L_4$ are given as
\begin{eqnarray}
X(t)&=&10^{-5}\left(4.28\,\cos{0.402\,t}-26.34\,\sin{0.402\,t}+\right.\nonumber\\&&\left.3.28\,\cos{0.915\,t}-11.86\,\sin{0.915\,t}\right),\end{eqnarray}\begin{eqnarray}
Y(t)&=&10^{-5}\left(7.02\,\cos{0.402\,t}-15.82\,\sin{0.402\,t}+\right.\nonumber\\&&\left.5.92\,\cos{0.915\,t}-6.89\,\sin{0.915\,t}\right),
\end{eqnarray}
which is composition of two periodic orbits with periods $T_1=\frac{1}{0.473}$ and $T_2=\frac{1}{0.875}$, respectively.
\section{Conclusion}
\label{sec:conclusion}

We have considered generalized restricted three body problem in which bigger primary is radiating-oblate spheroid, smaller primary is oblate body and a disc is rotating about the common center of mass of the system. Albedo effect of secondary has been analyzed in addition to the effect of radiation pressure, oblateness and disc in the context of zero velocity curves, existence of equilibrium points and their linear stability. Zero velocity curves for the Proxima Centauri system with dust belt and Sun-Saturn system with Kuiper belt are computed and analyzed with the help of Table \ref{tabzvc} and Figs. \ref{fig:zvca}-\ref{fig:zvcc}.
  It is found that
 on increasing the value of mass parameter $\mu$, the value of $C_j$ increases, which results an expansion in  the prohibited region. A similar nature have also seen for  Sun-Mars sytem with asteroid belt but due to minimize the length of paper, it is not presented. The equilibrium points of different planetary system such as Proxima Centauri system with dust belt, Sun-Mars-asteroid belt and Sun-Saturn-Kuiper belt are obtained and displayed in Table \ref{tab1} as well as in Figs.  \ref{fig:onea}-\ref{fig:onec}. The effects of perturbing parameters $q$, $Q_A$, $A_{21}$, $A_{22}$ and $M_{d}$ are analyzed for Proxima Centauri system with dust belt and it is noticed that oblateness of secondary have a remarkable effect on the positions of collinear equilibrium points $L_{1},\,L_{2},\,L_{3}$, whereas primary has less effect. On the other hand, effect of oblateness of the primaries is less on the positions of non-collinear equilibrium points $L_{4,5}$. The effect of radiation pressure and albedo on the positions of $L_{2,3}$ are normal, whereas in case of $L_1$ effect of albedo is more significant in comparison to that of radiation pressure. However, in case of $L_{4,5}$ both radiation pressure and albedo effects are significant.  It is also found that effect of the disc on the position of the equilibrium point is considerable (Table \ref{tab2} and \ref{tab3}). Linear stability of  collinear equilibrium points are analyzed with respect to mass parameter $\mu$ other remaining  perturbation parameters $q$, $Q_A$, $A_{21}$, $A_{22}$ and $M_d$ and it is found that $L_{1,2,3}$ are unstable in all case. However, range of linear stability of non-collinear equilibrium point contracted to  $0<\mu<\mu_{c}=0.0183124$ for $L_4\,(\,0.4934156,\,0.8620511)$ and  $0.0212928=\mu_0<\mu<\mu_{c}=0.042617$ for $L_{4}\,(0.442536,\,0.848967)$ (Fig. \ref{rootL4a}-\ref{rootL4b}). Variation of linear stability with respect to the parameters $q$, $Q_A$, $A_{21}$, $A_{22}$ and $M_d$ are also obtained (Fig \ref{rootL4b}).
Also, it is observed that the critical mass ratio $\mu_c$ increases with the increase in the values of mass reduction factor of first primary, oblateness of both the primaries and mass of the disc however, it decreases with the increment in albedo parameter of second primary (Table \ref{tab4}).  Finally, it is concluded that the effects of radiation pressure force, albedo force, oblateness and  the disc play a significant role in designing of the trajectories in the vicinity of equilibrium points \citep{lo1997libration,lo1998genesis,gomez1998station}. These results are useful to obtain more accurate results in other generalized problem of few body system. These results are limited to the regular symmetric disc and radiation pressure force, later it will be extended to P-R drag, solar wind drag etc.

\section*{Acknowledgments}
This work is partially supported by the University Grant Commission, India through the UGC start-up research grant No.-F.30-356/2017(BSR) and UGC-JRF Ref. No.-21/06/2016(i)EU-V, respectively. Some of the references used in this article are collected from the Library of Inter-University Center of Astronomy and Astrophysics (IUCAA), Pune (India).




\bibliographystyle{mnras}
\bibliography{RKSU_R1} 





\bsp	
\label{lastpage}
\end{document}